\documentclass[3p,times]{elsarticle}

\usepackage{ecrc}

\pdfmapfile{+txfonts.map}

\usepackage{epstopdf}

\volume{00}

\firstpage{1}

\journalname{Nuclear Physics A}

\runauth{Joseph I. Kapusta}

\jid{nupha}

\jnltitlelogo{Nuclear Physics A}

\usepackage{graphicx}
\usepackage{amsmath,amssymb}

\newcommand{\be}{\begin{equation}}
\newcommand{\ee}{\end{equation}}
\newcommand{\ba}{\begin{eqnarray}}
\newcommand{\ea}{\end{eqnarray}}
\newcommand{\bd}{\begin{displaymath}}
\newcommand{\ed}{\end{displaymath}}

\begin{document}

\begin{frontmatter}



\title{Early Time Dynamics of Gluon Fields in High Energy Nuclear Collisions}

\author{Joseph I. Kapusta}
\address{School of Physics and Astronomy, University of Minnesota,
Minneapolis MN 55455, USA}
\author{Guangyao Chen}
\address{Department of Physics and Astronomy, Iowa State University, 
Ames IA 50011, USA}
\author{Rainer J. Fries}
\address{Cyclotron Institute and Department of Physics and Astronomy,
Texas A\&M University, College Station TX 77843, USA}
\author{Yang Li}
\address{Department of Mathematics and Statistics, University of Minnesota
  -- Duluth,
Duluth MN 55812, USA}




\begin{abstract}
Nuclei colliding at very high energy create a strong, quasi-classical gluon field during the initial phase of their interaction. We present an analytic calculation of the initial space-time evolution of this field in the limit of very high energies using a formal recursive solution of the Yang-Mills equations. We provide analytic expressions for the initial chromo-electric and chromo-magnetic fields and for their energy-momentum tensor. In particular, we discuss event-averaged results for energy density and energy flow as well as for longitudinal and transverse pressure of this system. Our results are generally applicable if $\tau < 1/Q_s$. The transverse energy flow of the gluon field exhibits hydrodynamic-like contributions that follow transverse gradients of the energy density. In addition, a rapidity-odd energy flow also emerges from the non-abelian analog of Gauss' Law and generates non-vanishing angular momentum of the field. We will discuss the space-time picture that emerges from our analysis and its implications for observables in heavy ion collisions.
\end{abstract}


\end{frontmatter}



\section{Introduction}

Relativistic viscous fluid dynamics provides a very good description of particle distributions in high energy heavy ion collisions.  The initial conditions are usually taken at a proper time between 0.15 and 1.0 fm/c, with the initial energy or entropy density chosen to reproduce the total number of charged particles per unit rapidity at central rapidities.  The most promising approach for understanding the initial conditions is the color glass condensate.  The nuclei are modeled as a collection of color charges before the collision, leading to a classical gluon field immediately after the collision.  This field will soon decay into a thermalized plasma of quarks and gluons.  Our goal is to use the McLerran-Venugopalan model \cite{McLerran:1993ka,McLerran:1993ni,Gelis:2010nm} with boost invariance to carry out a Taylor series expansion in proper time in order to calculate the classical gluon field.  The color charge densities in the colliding nuclei are allowed to vary in the transverse directions, which leads to interesting and nontrivial energy and momentum flow fields.  Details of our calculations may be found in Ref. \cite{Fries:2005yc,Chen:2013ksa,us}.  

\section{Small Proper Time Expansion}

In the color glass condensate picture the color charges in the nuclei are essentially frozen during the very short time that it takes for the nuclei to pass through each other.  These charges are the source for the initial fields at $\tau=0$ which provide the initial conditions for solving the classical Yang-Mills equations.  Along the forward light cone
\begin{align}
  A_\perp^i (\tau=0,\vec x_\perp) &= A_1^i (\vec x_\perp) + A_2^i
  (\vec x_\perp),
  \label{eq:bc_boost1}\\
  A (\tau=0,\vec x_\perp) &= -\frac{ig}{2} \left[ A_1^i
  (\vec x_\perp),A_2^i (\vec x_\perp) \right].
  \label{eq:bc_boost2}
\end{align}
Here the index $i=1,2$ refers to one of the transverse ($\perp$) directions.  The Yang-Mills equations are
\begin{align}
  & \frac{1}{\tau}\frac{\partial}{\partial\tau}\frac{1}{\tau}
  \frac{\partial}{\partial\tau}
  \tau^2 A  - \left[ D^i ,\left[ D^i,A\right] \right] = 0 \, ,
  \label{eq:eom_noboost1}\\
  & ig\tau \left[ A, \frac{\partial}{\partial\tau}A \right]
  - \frac{1}{\tau} \left[ D^i, \frac{\partial}{\partial\tau}A_\perp^i \right]
  = 0 \, ,
  \label{eq:eom_noboost2}\\
  & \frac{1}{\tau}\frac{\partial}{\partial\tau}
  \tau\frac{\partial}{\partial\tau} A_\perp^i - ig\tau^2 \left[
  A, \left[ D^i,A \right]\right] - \left[ D^j, F^{ji} \right] = 0 \, .
  \label{eq:eom_noboost3}
\end{align}
In the chosen gauge the components of the field strength tensor are
\begin{align}
  F^{+-} &= -\frac{1}{\tau} \frac{\partial}{\partial \tau} \tau^2 A \,, \\
  F^{i\pm} &= -x^\pm \left( \frac{1}{\tau} \frac{\partial}{\partial\tau}
  A_\perp^i \mp [D^i,A] \right) \,, \\
  F^{ij} &= \partial^i A_\perp^j - \partial^j A_\perp^i -ig[A_\perp^i,
  A_\perp^j] \,.
\end{align}

We expand the gluon field in the following power series
\begin{align}
A(\tau,\vec x_\perp) = \sum_{n=0}^\infty \tau^n
A_{(n)}(\vec x_\perp) \;\;\; {\rm and} \;\;\;
A_\perp^i(\tau,\vec x_\perp) = \sum_{n=0}^\infty \tau^n
A_{\perp(n)}^i (\vec x_\perp) \, .
\end{align}
It turns out that all coefficients of odd powers vanish.  One finds the recursion relations for even $n$, $n>1$, to be
\begin{align}
A_{(n)} &= \frac{1}{n(n+2)} \sum_{k+l+m=n-2} \left[ D^i_{(k)}, \left[ D^i_{(l)}, A_{(m)} \right] \right] \,, \\
A^i_{\perp(n)} &= \frac{1}{n^2}\left( \sum_{k+l=n-2}
\left[ D^j_{(k)}, F^{ji}_{(l)} \right] +  ig \sum_{k+l+m=n-4} \left[ A_{(k)}, [ D^i_{(l)},A_{(m)} ] \right] \right)  \, .
\end{align}
Once solved to a given order in $\tau$ one can either consider heavy ion collisions on an event by event basis, in which case the color charge densities $\rho_1$ and $\rho_2$ of the two nuclei are chosen from Gaussian distributions, or one can average over collisions via
\begin{equation}
  \langle O \rangle_{\rho_1,\rho_2} = \int d[\rho_1]d[\rho_2] O(\rho_1,\rho_2)  w(\rho_1) w(\rho_2) \, .
\end{equation}
The weight functions $w$ are Gaussians with widths given by the average local charge densities squared, $\mu_1$ and $\mu_2$.

Several scales now enter the problem.  There is the local color charge squared per area $\mu$ and closely related to it is the so-called saturation scale $Q_s^2 \sim g^4 \mu$.  One also needs an IR regulator $\hat{m}$, which one can think of as a transverse gluon mass on the order of $\Lambda_{\rm QCD}$, and a transverse UV regulator $Q$, which is the scale separating the soft physics represented by the classical gluon fields and the hard physics represented by jets and mini-jets.  When the contribution to the energy momentum tensor from jets and mini-jets is included, which we are not doing here, the sum total should be relatively insensitive to the prescise choice of $Q$.

\section{Results}

The initial energy density at $Q \tau \ll 1$, neglecting transverse gradients, is
\begin{equation}
 \varepsilon_0 (\vec x_\perp) = 
\frac{2 \pi N_c \alpha_s^3}{ N_c^2-1 } \mu_1 (\vec x_\perp) \mu_2 (\vec x_\perp)
\ln \left(\frac{Q_1^2}{\hat m^2}\right) \ln \left(\frac{Q_2^2}{\hat m^2}\right) \,.
\label{ini_den}
\end{equation}
The appearance of $\alpha_s$ to the third power can be viewed as arising from the amplitude for emission of gluons from each nuclei followed by their fusion into one gluon.  One would expect that quantum corrections would replace one power of $\alpha_s$ with a renormalization group running coupling $\alpha_s(Q_1^2)$, another by $\alpha_s(Q_2^2)$, and the last one at the triple gluon vertex by $\alpha_s(Q^2)$.  Then, using $\alpha_s(M^2) = 1/\left[\beta_2 \ln (M^2/\Lambda^2_{\rm QCD})\right]$ with $\beta_2 = (11N_c - 2N_f)/12\pi$ we would get
\begin{equation}
 \varepsilon_0 (\vec x_\perp) \approx 
\frac{2\pi  N_c \alpha_s(Q^2)}{ \beta_2^2 (N_c^2-1) } \,  
\mu_1 (\vec x_\perp) \mu_2 (\vec x_\perp) \, .
\end{equation}
It would appear that scale dependences are weaker once quantum corrections are established. Of course, the functions $\mu_i(\vec x_\perp)$ also depend to some degree on the scales. 

\begin{figure}[h]
\begin{center}
\includegraphics*[width=16.00cm]{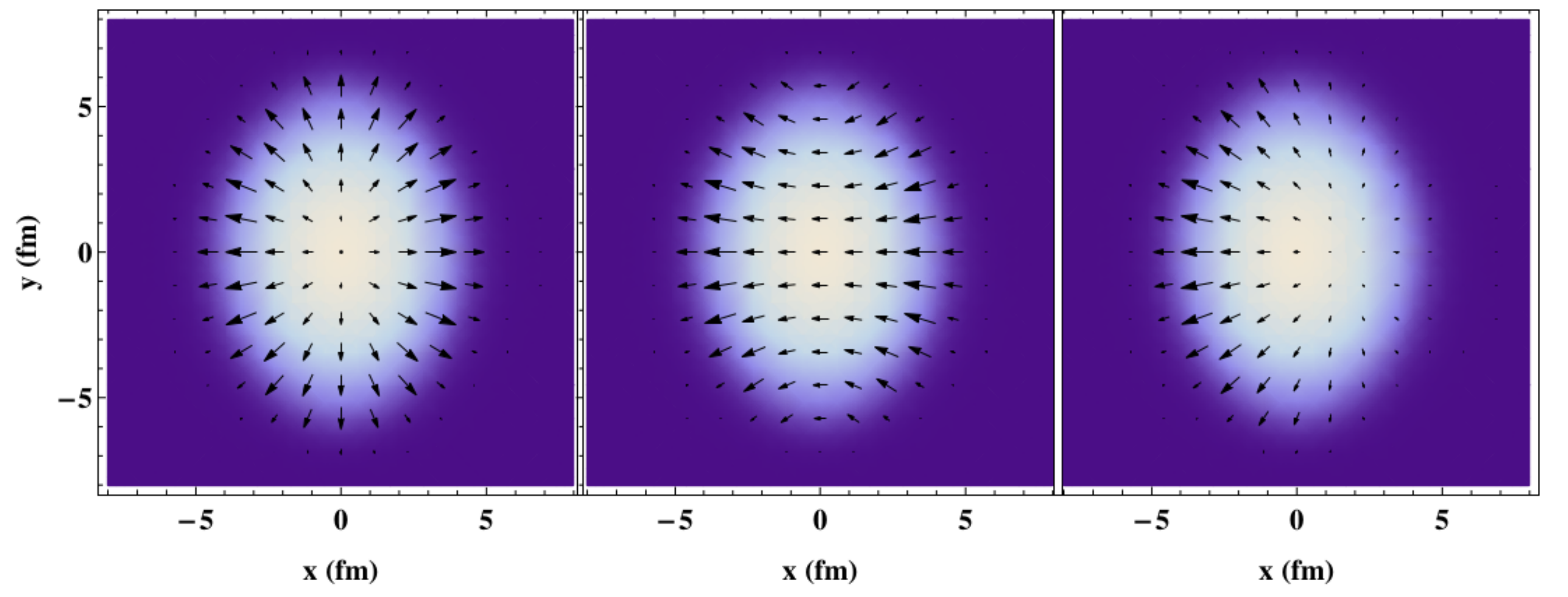}
\caption{Different flow fields (arrows) and initial energy density $\varepsilon_0$
     (shading) for Pb+Pb collisions at impact parameter $b=6$ fm in the 
     $x-y$-plane. Lighter regions correspond to higher densities.  The nucleus centered at $x=3$ fm travels in the positive 
     $\eta$-direction.  Left panel: Rapidity even contribution $\alpha^i$. Middle panel: Rapidity odd contribution $\beta^i$.
     Right panel: Full transverse Poynting vector $T^{0i}$ at $\eta=1$.}
\label{Fig1}
\end{center}
\end{figure}

As time increases gradients in the color charge densities lead to transverse flow of energy and momentum.  Up to order $\tau^2$ the transverse energy flow is expressed in terms of rapidity even and odd terms as
\begin{equation}
  T^{0i}_{\mathrm{even}} =
  \frac{\tau}{2} \, \alpha^i \left( 1 -\frac{1}{2a} 
  (Q\tau)^2  \right)  \cosh\eta \;\;\; {\rm and} \;\;\;
  T^{0i}_{\mathrm{odd}} =
  \frac{\tau}{2} \, \beta^i \left( 1- \frac{9}{16a} 
  (Q\tau)^2  \right) \sinh\eta \, .
\end{equation}
(The functions $\alpha^i$ and $\beta^i$ are proportional to the gradients of $\mu_1$ and $\mu_2$.)  These terms are shown in Fig. \ref{Fig1}.  Such initial flows are generally not included in current viscous fluid models.

\begin{figure}[ht] 
\begin{center}
\includegraphics[width=12.00cm]{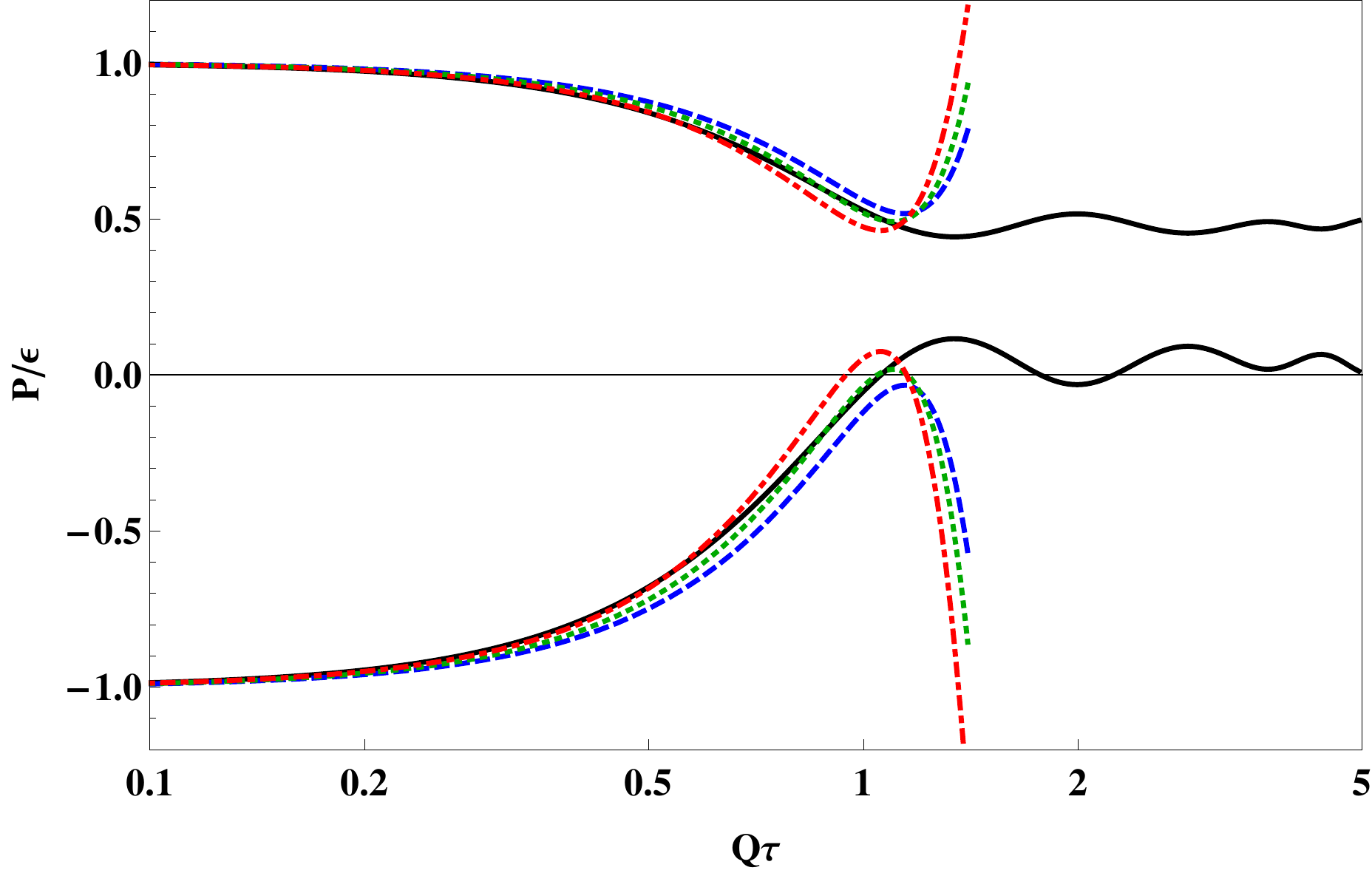}
\caption{Evolution of the ratios of the transverse (upper
    curves) and longitudinal (lower curves) pressure over energy density for
    the classical gluon field at fourth order accuracy in time, compared to
    the numerical result from \cite{Gelis:2013rba} at leading order 
    for $g=0.5$ (solid lines). Values of $\ln(Q^2/\hat{m}^2)$ = 0.8, 0.9, and 1.0 (dash-dotted, dotted and dashed
    lines respectively) are indicated.}
\label{Fig2}
\end{center}
\end{figure}

Next consider the case of identical slab on slab collisions.  There are no transverse gradients.  The longitudinal pressure $P_L$ is defined by $T^{33}$ and the transverse pressure $P_T$ by $T^{ii}$ with $i$ either 1 or 2.  The ratio of these pressures to the energy density is shown in Fig. \ref{Fig2} up to and including terms of order $\tau^4$.  We can roughly compare our results with those of Ref. \cite{Gelis:2013rba}, who performed a real-time lattice simulation for colliding slabs using the gauge group $SU(2)$ and with $g=2$.  Up to times of order $1/Q$ the agreement between the numerical results and our analytical calculations are quite reasonable.  What is most interesting is that the system is attempting to become isotropic with a positive $P_L$.  One can show from our expression for the energy-momentum tensor that if $P_L = P_T \equiv P$ could be achieved then necessarily $P = \varepsilon/3$ which is the equation of state for a weakly interacting gas of massless particles.   Of course, what is missing now is how the classical field decays into quarks and gluons and thermalizes.

\section{Conclusion}

We have worked out analytic solutions of the Yang-Mills equations for two nuclei with random color charges colliding on the light cone.  Using a recursive solution we computed the early time gluon field and energy-momentum tensor in a near-field approximation. We have also calculated expectation values for the energy-momentum tensor when many events are averaged. Our calculation generalizes the McLerran-Venugopalan model to allow small but non-vanishing gradients in the average color charge in the transverse plane. This permitted us to discuss flow phenomena in the averaged events.  We find that this approximation gives acceptable results roughly up to a time given by $Q \tau \approx 1$. This coincides with the time at which the entire classical field approximation probably starts to break down.

The transverse flow of energy starts out from zero, initially grows linearly in time, and might reach significant values at the surface of the fireball. The time evolution of transverse and longitudinal pressure matches well with numerical results available in the literature up to $Q \tau \approx 1$.  Besides the usual radial and elliptic flow, a rapidity-odd flow emerges. We suggest that this energy flow of the glasma could contribute to directed flow measured at RHIC and LHC. It carries angular momentum which rotates the fireball. The characteristic glasma flow pattern could potentially lead to another signature for color glass dynamics in high energy collisions.

Future work would include calculating explicitly the coefficients of higher powers in $\tau$ and optimistically matching the results to a 3+1 dimensional viscous hydrodynamic code.  Conversion of the classical gluon field to a thermalized quark gluon plasma remains an open problem.

This work was supported by the Office of Science, U. S. Department of Energy, and by the U. S. National Science Foundation.






\begin{thebibliography}{00}


\bibitem{McLerran:1993ka}
L. D. McLerran and R. Venugopalan, Phys. Rev.  D {\bf 49}, 3352 (1994).

\bibitem{McLerran:1993ni}
L. D. McLerran and R. Venugopalan, Phys. Rev. D {\bf 49}, 2233 (1994).

\bibitem{Gelis:2010nm}
F. Gelis, E. Iancu, J. Jalilian-Marian, and R. Venugopalan, Ann. Rev. Nucl. Part. Sci.  {\bf 60}, 463 (2010).

\bibitem{Fries:2005yc} 
R. J. Fries, J. I. Kapusta, and Y. Li,  Nucl. Phys. A {\bf 774}, 861 (2006).

\bibitem{Chen:2013ksa}
G. Chen and R. J. Fries,  Phys. Lett. B {\bf 723}, 417 (2013).

\bibitem{us}
G. Chen, R. J. Fries, J. I. Kapusta, and Y. Li, Phys. Rev. C {\bf 92}, 064912 (2015).

\bibitem{Gelis:2013rba}
T. Epelbaum and F. Gelis, Phys. Rev. Lett. {\bf 111}, 232301 (2013).


\end{thebibliography}



\end{document}